\documentclass{article}

\usepackage[numbers,compress]{natbib}
\usepackage[dblblindworkshop, final]{neurips_2025}
\workshoptitle{}

\usepackage[utf8]{inputenc} 
\usepackage[T1]{fontenc}    
\usepackage{hyperref}       
\usepackage{url}            
\usepackage{booktabs}       
\usepackage{amsfonts}       
\usepackage{nicefrac}       
\usepackage{microtype}      
\usepackage{xcolor}         
\usepackage{comment}
\usepackage{amsmath}
\usepackage[pdftex]{graphicx}

\title{When Less is More: Approximating the Quantum Geometric Tensor with Block Structures}

\author{%
  Ahmedeo Shokry$^{1,2,3,4}$ \quad
  Alessandro Santini$^{1,2,3,4}$ \quad
  Filippo Vicentini$^{1,2,3,4}$ \\[4pt]
  $^1$CPHT, CNRS, École Polytechnique, Institut Polytechnique de Paris, 91120 Palaiseau, France \\
  $^2$Collège de France, Université PSL, 11 place Marcelin Berthelot, 75005 Paris, France \\
  $^3$Inria Paris–Saclay, Bâtiment Alan Turing, 1 rue Honoré d’Estienne d’Orves, 91120 Palaiseau, France \\
  $^4$LIX, CNRS, École Polytechnique, Institut Polytechnique de Paris, 91120 Palaiseau, France \\[4pt]
  \texttt{\{ahmedeo.shokry, alessandro.santini, filippo.vicentini\}@polytechnique.edu}
}
\hypersetup{colorlinks=true, citecolor=blue, linkcolor=blue, urlcolor=blue}
\begin{document}

\maketitle

\begin{abstract}
The natural gradient is central in neural quantum states optimizations but it is limited by the cost of computing and inverting the quantum geometric tensor, the quantum analogue of the Fisher information matrix. We \textcolor{black}{study} a block-diagonal quantum geometric tensor that partitions the metric by network layers, analogous to block-structured Fisher methods such as K-FAC. This layer-wise approximation preserves essential curvature while removing noisy cross-layer correlations, improving conditioning and scalability. Experiments on Heisenberg and frustrated $J_1$–$J_2$ models show faster convergence, lower energy, and improved stability.
\end{abstract}

\section{Introduction}

A fundamental problem in quantum many-body physics is to determine the ground-state wavefunction that minimizes the system’s energy. This is equivalent to solving an eigenvalue problem for an exponentially large Hamiltonian matrix, a task far beyond the reach of exact diagonalization. In practice, variational Monte Carlo (VMC)~\citep{Becca_Sorella_2017} reformulates this as a stochastic optimization problem: the loss function (the energy) and its gradients are estimated through Monte Carlo sampling. Samples are drawn from the square of the current parametrized wavefunction, either through ancestral sampling or Markov chain Monte Carlo (MCMC).

Neural-network quantum states (NQS)~\citep{carleo2017troyer, Vicentini2021} address the core challenge of VMC: representing and optimizing a high-dimensional, complex-valued wavefunction. 
NQS avoid the exponential memory complexity of quantum mechanics by expressing the wavefunction as a neural network that takes the system configurations as input and outputs its complex amplitude. Within this framework, state-of-the-art accuracy has been demonstrated on challenging benchmarks including the Fermi–Hubbard model~\citep{arovas2022, qin2022, hfds, Chen:2025} and the frustrated $J_1$–$J_2$ magnet \cite{nutakki2025design,Chen2024, roth2023a}. Extensions to variational dynamics~\cite{schmitt_quantum_2022,gravina_neural_2024,Sinibaldi2024dynamics}, excited states~\cite{Pfau2024Science} and subspaces exist~\cite{Kahn2025Subspace,Hendry2025Grassman}.

However, standard first-order optimizers such as Adam~\cite{Kingma2014AdamAM} often plateau before reaching the accuracies required in physics. 
In the usual formulation, natural-gradient updates are employed to exploit the underlying geometry of the parameter space, an approach known in physics as \emph{stochastic reconfiguration} (SR)~\cite{sorella}, and which has rapid convergence guarantees~\citep{Goldshlager2025FastNGD}.
In the language of machine learning (ML), this corresponds to preconditioning the energy gradient with the inverse of the Fisher information matrix (FIM), known in the literature of complex wave-functions as the \emph{quantum geometric tensor} (QGT). 
As the FIM is a $n_p\times n_p$ matrix, where $n_p$ is the number of neural network parameters, inverting it quickly becomes untractable.
As it's usually ill-conditioned, LU decompositions are unreliable and it is often required to perform a full diagonalization with a complexity scaling as $\mathcal{O}(n_p^3)$ and which cannot be easily parallelized across multiple devices. 
Recent sample-space reformulations \citep{Chen2024,Rende2024} mitigate this by inverting the neural tangent kernel (NTK)-like matrix whose size depends on the number of Monte Carlo samples rather than the parameters~\citep{ren2019efficientsubsampledgaussnewtonnatural}.

In this paper, inspired by FIM block diagonal approaches such as KFAC and mini block Fisher (MBF) ~\citep{inproceedingsLecun,pmlr-v28-schaul13,NIPS2007_9f61408e,10.1093/imaiai/iav006} we benchmark the block-layer QGT approximation~\citep{Nys_2024} against the full QGT and NTK. In particular, we apply the technique to (i) the Heisenberg chain and (ii) the highly frustrated 2D $J_1\text{–}J_2$ lattice. We find that employing the block approximation we increase the effective rank of the preconditioner, mitigate ill-conditioning~\citep{Dash2025}, achieve better convergence, and yield faster training. 

\section{Natural Gradient Descent for NQS and Block-Layer QGT}

For lattice systems with spin degrees of freedom, the neural network representing the quantum system takes as input a configuration $\mathbf{x} \in \{-1, +1\}^N$ and outputs a complex amplitude $\psi_\theta(\mathbf{x})$, defining a variational wavefunction parameterized by the weights $\theta$. The parameters are optimized to minimize the variational energy, i.e., the expectation value of the Hamiltonian
\begin{equation}
E = \frac{\langle \psi_\theta | H | \psi_\theta \rangle}{\langle \psi_\theta | \psi_\theta \rangle} = \mathbb{E}_{\mathbf{x} \sim p_{\theta}(\mathbf{x})}\bigg[\frac{\langle \mathbf{x}|H|\psi_{\theta} \rangle}{\langle \mathbf{x}|\psi_{\theta} \rangle}\bigg],
\end{equation}
where expectation values are estimated via Monte Carlo sampling from the Born distribution $p_{\theta}(\mathbf{x}) = |\psi_\theta(\mathbf{x})|^2$. As the Hilbert space grows exponentially with system size, exact evaluation becomes infeasible, making Monte Carlo sampling essential for NQS training. 

Training, however, remains challenging: Monte Carlo estimates introduce noise and the energy landscape is highly nonconvex.  Methods like stochastic gradient descent and Adam often fail to converge, motivating geometry-aware approaches like natural gradient descent (NGD)~\citep{amari}.
NGD follows updates determined by the solution of the linear system $S \cdot \delta\theta=-F$, where $F_i = \partial_{\theta_i} E$ (the stochastic energy gradient~\cite{2507.05352}) and $S$ (the QGT) are estimated from the same Monte Carlo samples. 
For a variational wavefunction $\psi_\theta$, the QGT is defined as
\begin{equation}
S_{ij} = \frac{\langle \partial_{\theta_i} \psi_{\theta} | \partial_{\theta_j} \psi_{\theta} \rangle }{\langle \psi_{\theta} | \psi_{\theta} \rangle} - \frac{\langle \partial_{\theta_i} \psi_{\theta}| \psi_{\theta}\rangle}{\langle \psi_{\theta} | \psi_{\theta} \rangle} \frac{\langle \psi_{\theta}| \partial_{\theta_j}\psi_{\theta}\rangle}{\langle \psi_{\theta} | \psi_{\theta} \rangle} = \mathbb{E}_{\mathbf{x}\sim p_{\theta}(\mathbf{x})}\big[\Delta O^{\dagger}_i(\mathbf{x})\Delta O_j(\mathbf{x}) \big],
\end{equation}
with $\Delta O_i(\mathbf{x}) = O_i(\mathbf{x})-\langle O_i \rangle$ and $O_i(\mathbf{x}) = \partial_{\theta_i} \log \psi_{\theta}(\mathbf{x})$.
Formally, the natural gradient update is $\delta \theta = -\eta \cdot S^{-1} F$,
where $\eta$ is a learning rate and $S^{-1}$ is a pseudo-inverse.
\paragraph{Block-layer QGT.} Consider a variational wavefunction $\psi_{\theta}(\mathbf{x})$,  
whose parameters $\theta$ are partitioned into $L$ disjoint groups corresponding to distinct modules: a patching and embedding layer $L_{\text{emb}}$, transformer encoder layers $\{ L_{\text{E}(1)} \ ... \ L_{\text{E}(N)} \}$, and a final output layer $L_{\text{out}}$. 
Denoting the parameters of layer $l$ as $\theta_l$, we write $\theta = (\theta_{L_{\text{emb}}}, \theta_{L_{\text{E}(1)}}, \ ... \ \theta_{L_{\text{E}(N)}}, 
\theta_{L_{\text{out}}})$.
We construct a block-diagonal approximation of the QGT aligned with the network decomposition
\begin{equation}
S \approx S^\text{block} = {\rm diag}(S_{L_{\text{emb}}}, S_{L_{\text{E}(1)}},\ \cdots,  S_{L_{\text{out}}} ),
\qquad S_l = \langle O_l^\dagger O_l \rangle - \langle O_l^\dagger \rangle \langle O_l \rangle,
\end{equation}
where $O_l$ is the Jacobian of the log-wavefunction with respect to layer parameters $\theta_l$.
Each block $S_l$ is inverted independently, using a uniform diagonal shift $\lambda$ to stabilize the inversion, $\tilde{S}_l = S_l + \lambda I_l,$ and $ \delta \theta_l = -\eta \tilde{S}_l^{-1} F_l$.
The full parameter update is $\delta\theta = (\delta\theta_{L_{\text{emb}}}, \delta\theta_{L_{\text{E}(1)}}, \ ... \ \delta\theta_{L_{\text{E}(N)}}, 
\delta\theta_{L_{\text{out}}})$.
The block-layer QGT method differs from prior approaches in several ways. (i) Unlike KFAC~\citep{martens15,grosse2016}, which factorizes each layer’s FIM block as a Kronecker product, we treat each block as a full matrix, preserving all intra-layer correlations. (ii) Compared to MBF~\citep{Bahamou2022AMF}, which subdivides layers into smaller blocks, our decomposition follows the natural structure of the network modules (i.e., embedding, encoders, output), which has a straightforward extension to deep transformers.
\section{Numerical Results}
To evaluate the quality of the block-diagonal approximation, we consider metrics defined on the corresponding inverse operators. 
Let $A$ and $B$ denote the inverses of the block-layer QGT and exact QGT, respectively. Since the QGT defines a metric, approximations may distort geometric orientation (eigenvectors) or spectral scaling (eigenvalues). We therefore use complementary diagnostics probing both alignment and spectral fidelity.
\begin{equation}
\begin{alignedat}{2}
\textbf{(a)}\quad 
&\mathrm{F}(A,B) = \frac{\mathrm{Tr}[A^\dagger B]}{\|A\|_F\,\|B\|_F}, 
\quad &&\textbf{(b)}\quad\epsilon_F(A,B) = \frac{\|A-B\|_F}{\|B\|_F}, \\[4pt]
\textbf{(c)}\quad 
&\kappa(A) = \frac{\lambda_{\max}(A)}{\lambda_{\min}(A)}, 
\quad &&\textbf{(d)}\quad r_\lambda(A,B) = \mathrm{corr}\!\Big(
        \{\lambda_i^{(A)}\}_{\!\downarrow},
        \{\lambda_i^{(B)}\}_{\!\downarrow}
      \Big).
\end{alignedat}
\end{equation}
\textbf{(a) Frobenius overlap:} Cosine of the angle between \(A\) and \(B\) in matrix space. 
\(\mathrm{F}(A,B)\!\approx\!1\) gives collinear natural-gradient updates for any force \(F\). 
\textbf{(b) Frobenius relative error:} 
Average relative deviation between \(A\) and \(B\) over all matrix entries. \textbf{(c) Condition number:}  
Numerical conditioning of the natural gradient. \textbf{(d) Eigenvalue spectrum correlation:}  
Here \(\{\lambda_i\}_{\!\downarrow}\) are eigenvalues sorted in descending order.  
Values \(r_\lambda\!\approx\!1\) indicate that the approximation preserves the relative scaling of curvature modes, determining the effective step-size weighting in natural-gradient updates.

\paragraph{Heisenberg Chain.}
As a first benchmark, we consider the spin-$1/2$ Heisenberg model on a chain of length $L=16$ with periodic boundary conditions. The Hamiltonian reads $H =  \sum_{i=1}^{L} \mathbf{S}_i \cdot \mathbf{S}_{i+1}$, where $\mathbf{S}_i = (\sigma_i^x, \sigma_i^y, \sigma_i^z)/2$ are the spin-$1/2$ operators at site $i$. 
This system is small enough to be exactly solvable while still exhibiting nontrivial correlations, providing a clean benchmark for our block-layer QGT against full SR and exact results. 
We trained the same Vision-Transformer network~\citep{viterittiPRLViT} for all three settings, using a two-layer encoder. The optimized QGT was stored after 2000 epochs, along with the energy values during optimization and the exact infidelity between the states obtained with the sampled QGT and the exact QGT. Where the infidelity between two quantum states is defined as $\mathcal{I}(|\psi\rangle,|\phi\rangle)
= 1 - |\langle\psi|\phi\rangle|^2/(\langle\psi|\psi\rangle\,\langle\phi|\phi\rangle)$,
vanishing for identical states and reaching one for orthogonal ones; it is directly related to the Fubini–Study metric~\citep{fubini,study}.
\begin{figure}[h]
  \centering
  \includegraphics[width=1.\linewidth]{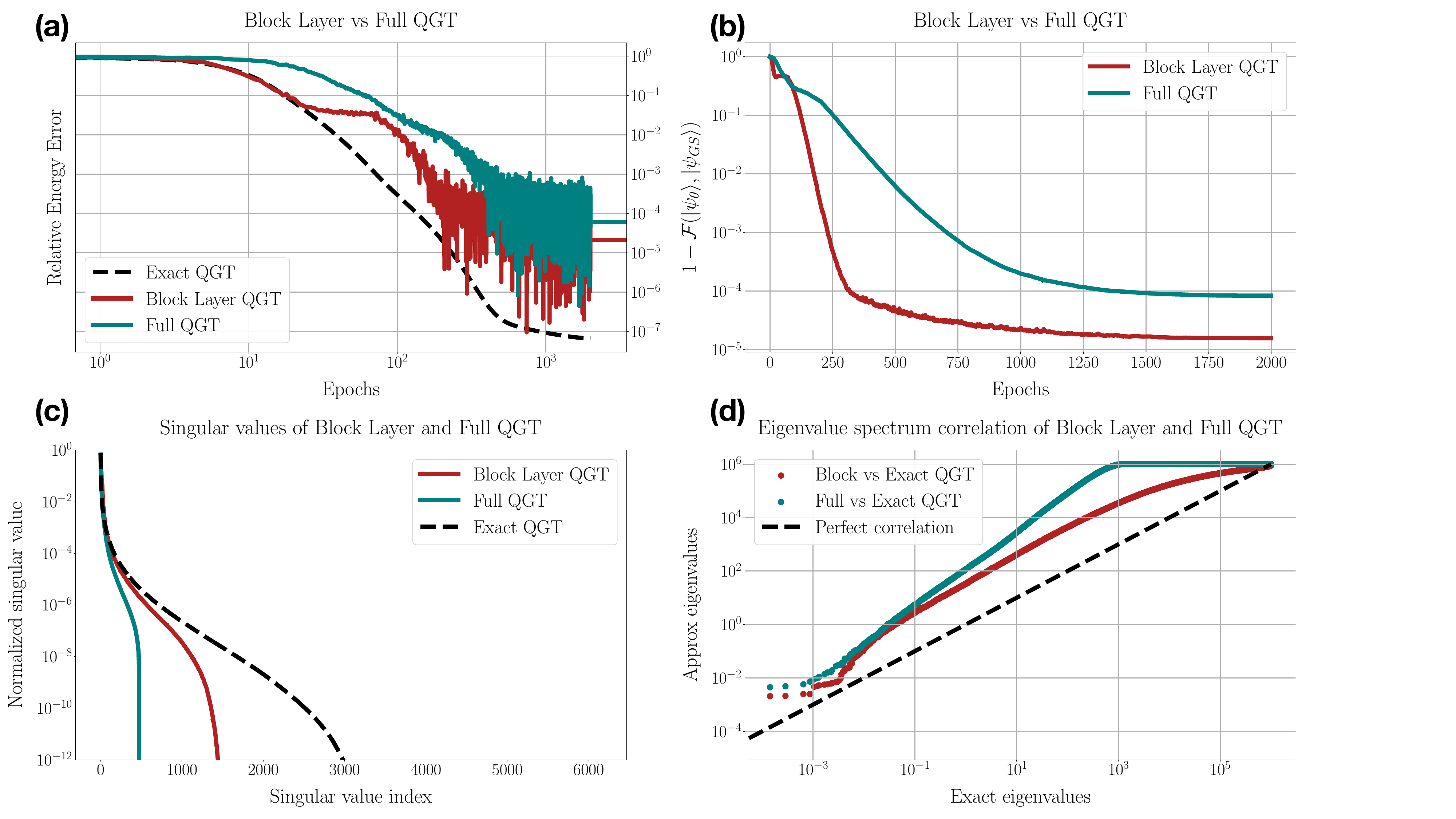}
  \caption{Comparison of block-layer and full QGT optimization for the $L=16$ Heisenberg chain. (a) Relative energy error; (b) infidelity to the exact ground state; (c) normalized QGT spectra at convergence; (d) correlation between approximate and exact QGT eigenvalues.}
\end{figure}
 \begin{table}[h!]
  \caption{Diagnostics comparing block-layer QGT and full sampled QGT against the exact QGT} 
  \label{tab:qgt-blockwise}
  \centering
  \begin{tabular}{lllll}
    \toprule        
    \cmidrule(r){1-2}
    \text{Metric} & \text{Block-layer vs Exact} & \text{Full vs Exact} \\
    \midrule
    Frobenius overlap $\uparrow$ & $0.656$ & \textbf{\boldmath$0.702$}  \\
    Frobenius relative error $\downarrow$ & \textbf{\boldmath$0.834$} & $0.864$  \\
    Condition number $\downarrow$ & \textbf{\boldmath $2.15 \cdot 10^{9}$} & $2.38\cdot10^{9}$  \\
    Eigenvalue spectral correlation $\uparrow$ & \textbf{\boldmath$0.979$} & $0.639$  \\ 
    \bottomrule
  \end{tabular}
\end{table}

Table~\ref{tab:qgt-blockwise} summarizes a comprehensive comparison between the block-layer and full QGT approximations against the exact QGT computed on the full Hilbert space. All metrics are evaluated on the inverse metric tensor. 
The results reveal a consistent pattern. Although the full QGT achieves a slightly higher Frobenius overlap with the exact inverse (0.702 vs. 0.656), the block-layer approximation better preserves the eigenvalue spectrum, with a correlation of 0.979 compared to 0.639 for the full QGT. This shows that the block formulation retains the relative weighting of geometric modes—crucial for natural-gradient updates, which depend on the eigenvalue structure of $S$. Across other metrics, both methods reach similar accuracy, but the block-layer QGT yields smaller Frobenius relative error (0.834 vs. 0.864) and smaller condition number ($2.15 \cdot 10^9$ vs $2.38 \cdot 10^9$).

This structural fidelity translates directly into optimization performance: the accurate eigenvalue ranking of the block-layer QGT produces correctly scaled parameter updates, faster convergence, and lower final infidelity. Furthermore, Table \ref{tab:measurements} shows faster convergence and reduced simulation time.
 \begin{table}[h!]
  \caption{Summary of measurements for the chain $L=16$ Heisenberg model.}
  \label{tab:measurements}
  \centering
  \begin{tabular}{lllll}
    \toprule        
    \cmidrule(r){1-2}
    Method & Energy $E$ & Energy Variance $\sigma^2$ & Infidelity $\mathcal{I}$ & Wall Time \\
    \midrule
    Block-layer QGT  & -28.5685(3) & 0.008 & $1.6 \cdot 10^{-05}$ & $\sim$ 25 min \\
    Full QGT   & -28.5674(6) &  0.026 & $8.4 \cdot 10^{-05}$ & $\sim$ 1 hour \\
    \bottomrule
  \end{tabular}
\end{table}
\paragraph{Frustrated $J_1\text{–}J_2$ model.} We consider the spin-$1/2$ $J_1\text{–}J_2$ Heisenberg model on a square lattice with periodic boundary conditions as a concrete benchmark of our method. The Hamiltonian reads $H = J_1 \sum_{\langle i,j \rangle} \textbf{S}_i \cdot \textbf{S}_j + J_2 \sum_{\langle \langle i,j \rangle \rangle } \textbf{S}_i \cdot \textbf{S}_j
$, where $\langle i,j \rangle$ and $\langle \langle i,j \rangle \rangle$ denote nearest- and next-nearest-neighbor pairs, respectively.  When the ratio $J_2 /J_1$ is in the range $\sim 0.4 - 0.6$, the system enters a highly frustrated regime where competing interactions prevent simple magnetic order and give rise to competing states with small energy gaps, making variational optimization challenging.
\begin{figure}[h]
  \centering
  \includegraphics[width=1.\linewidth]{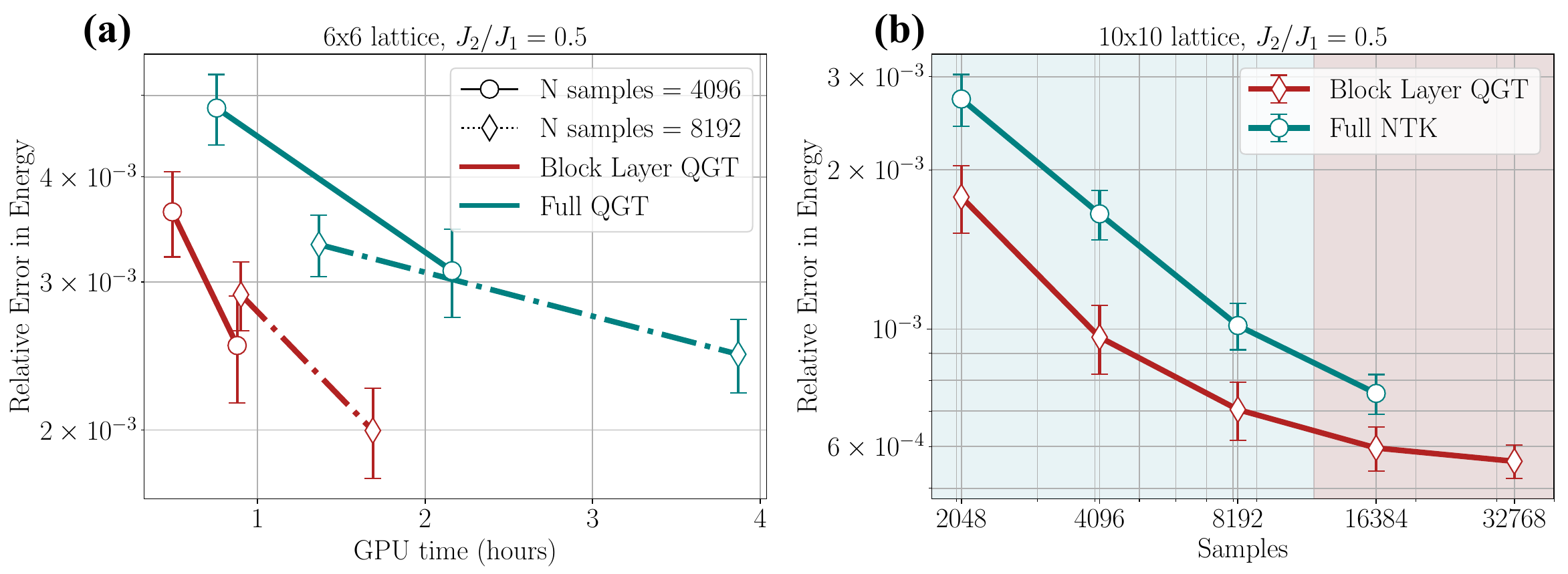}
  \caption{Comparison of block-layer and full QGT optimization in the $J_1\text{–}J_2$ model at $J_2/J_1=0.5$. (a): $6\times6$ lattice—energy error vs. GPU time. (b): $10\times10$ lattice—energy error vs. samples. Blue (red) shaded background, faster NTK (block-layer QGT) convergence.}
  \label{fig:lattices}
\end{figure}
On the $6\times 6$ lattice [Fig.~\ref{fig:lattices}(a)], we see that the block-layer QGT converges significantly more rapidly than the full QGT and remains stable even when only a modest number of Monte Carlo samples are available. On the larger $10\times 10$ lattice [Fig.~\ref{fig:lattices}(b)], this advantage becomes more pronounced: across all tested sample sizes, block-layer QGT consistently reaches lower relative energy errors, which we evaluate with respect to the best variational energy shown in~\citep{Chen2024}. This improvement stems from the spectral properties of the QGT: the block approximation suppresses noisy cross-layer correlations, mitigating near-degeneracies and improving conditioning. For comparison, the NTK approach is computationally cheaper and faster at small sample sizes but becomes unstable beyond $2^{14}$ samples. In contrast, the block-layer QGT remains stable, accurate, and competitively fast, striking a favorable balance between efficiency and robustness.
\section{Conclusions}
We observed that a block-layer-diagonal QGT is less rank-deficient than the standard estimator, better capturing its spectrum, while reducing the computational cost at large sample count.
This approach provides a scalable alternative to full-matrix natural gradients, achieving faster convergence, lower energies, and more stable training. 
Beyond VMC, block-structured curvature approximations may improve optimization in other differentiable scientific simulators where full-matrix natural gradients are prohibitive.

\begin{ack}
    Simulations were performed with custom-built extensions to NetKet~\cite{netket2:2019,vicentini2022netket}, and at times parallelized with mpi4JAX~\cite{mpi4jax:2021}. 
    This software is built on top of JAX \cite{jax2018github} and Flax \cite{flax2020github}. We acknowledge insightful discussion with J. Nys. 
    F.V. acknowledges support by the French Agence Nationale de la Recherche through the NDQM project ANR-23-CE30-0018.
    We acknowledge EuroHPC Joint Undertaking for awarding us access to Leonardo at CINECA, Italy through grant EHPC-AI-2024A05-006 .
\end{ack}

\appendix

\end{document}